\documentclass[doublespacing]{elsart}

\usepackage{graphicx}
\usepackage{array,tabularx}
\newcommand{\pt}{$p_T$}
\newcommand{\Ks}{$K^{0}_{S}$ }
\newcommand{\La}{$\Lambda$ }

\usepackage{epsfig}

 \usepackage{amssymb}

\usepackage{lineno}
\linenumbers
\begin{document}

\begin{frontmatter}

\title{Improving the $dE/dx$ calibration of the STAR TPC for the high-$p_T$ hadron identification}
\author [ustc,bnl] {Yichun Xu \corauthref{cor1}}\ead{xuychun@mail.ustc.edu.cn},
\author [uic] {Olga Barannikova},
\author [uw] {Hans Bichsel},
\author [ustc,lbl] {Xin Dong},
\author [bnl]  {Patricia Fachini},
\author [bnl] {Yuri Fisyak},
\author [mit] {Adam Kocoloski},
\author [vecc] {Bedanga Mohanty},
\author [purdue] {Pawan Netrakanti},
\author [bnl] {Lijuan Ruan},
\author [uic] {Maria Cristina Suarez},
\author [ustc,bnl] {Zebo Tang},
\author [bnl] {Gene van Buren},
\author [ustc,bnl] {Zhangbu Xu}
\corauth[cor1]{+1-631-344-7635}
\address [ustc] {Department of Modern Physics, University of Science and Technology of China, Hefei, Anhui, China, 230026}
\address [bnl]{Department of Physics, Brookhaven National Laboratory, Upton, NY 11973, USA}
\address[uw]{Nuclear Physics Laboratory, University of Washington, Box 354290, Seattle, WA 98195-4290, USA}
\address [lbl]{Lawrence Berkeley National Laboratory, 1 Cyclotron Road, Berkeley, CA 94720, USA}
\address [mit] {Massachusetts Institute of Technology, 77 Massachusetts Avenue,
Cambridge, MA 02139-4307, USA}
\address [vecc]{Variable Energy Cyclotron Centre, Kolkata 700064, India}
\address [purdue]{Purdue University, West Lafayette, Indiana 47907, USA}
\address [uic]{University of Illinois at Chicago, Department of Physics
845 W. Taylor St., M/C 273 Chicago, IL 60607-7059, USA }

\begin{abstract}
% Text of abstract

We derive a method to improve particle identification (PID) at high
transverse momentum ($p_T$) using the relativistic rise of the
ionization energy loss ($dE/dx$) when charged particles traverse the
Time Projection Chamber (TPC) at STAR. Electrons triggered and
identified by the Barrel Electro-Magnetic Calorimeter (BEMC), pure
protons(anti-protons) and pions from $\Lambda$ ($\overline{\Lambda}$),
and $K^{0}_{S}$ decays are used to obtain the $dE/dx$ value and its
width at given $\beta\gamma=p/m$. We found that the deviation of the
$dE/dx$ from the Bichsel function can be up to $0.4\sigma$ ($\sim3\%$)
in p+p collisions at $\sqrt{s_{NN}}=200$ GeV taken and subsequently
calibrated in year 2005. The deviation is approximately a function of
$\beta\gamma$ independent of particle species and can be described
with the function $f(x) = A+\frac{B}{C+x^{2}}$. The deviations
obtained with this method are used in the data sample from p+p
collision for physics analysis of identified hadron spectra and their
correlations up to transverse momentum of 15 GeV/$c$. The ratio of
$e^{-}/e^{+}$ (dominantly from $\gamma$-conversion) is also used to
correct for the residual momentum distortion in the STAR TPC.

\end{abstract}

\end{frontmatter}

% main text
\section{Introduction}
%physical goals(??)=>detector(TPC+EMC)

Identified hadron ($\pi^{\pm}$, $K^{\pm}$, $p(\overline{p}$))
spectra at high $p_T$ in p+p collisions provide a good test of
perturbative quantum chromodynamics (pQCD) \cite{pQCD} and a
baseline for studying color charge effect of parton energy loss in
heavy ion collisions \cite{starPIDpapers,starppPID}. Hadron
identification at high $p_T$ was achieved with the ionization energy
loss ($dE/dx$) at the relativistic rise using the STAR Time
Projection Chamber (TPC). The $dE/dx$ for a given charged-particle
track is calculated using a well-known method called "truncated
mean". The $dE/dx$ values of hits associated with the track have a
typical Landau tail. The hits with the top 30\% of high $dE/dx$
values are discarded and an average of $dE/dx$ value from the rest
of the hits is derived for that track~\cite{TPC}. The $dE/dx$ for a
given particle at low momentum decreases with increasing momentum to
reach a minimum ionization, then increases due to the relativistic
rise. For a minimum ionizing particle (MIP) the $dE/dx$ resolution
in the STAR TPC is 6-8\% for a track with the maximum of 45 sampled
$dE/dx$ points. The pions are well separated from the rest of the
particles ($e,K,p$) at $p_T$ of 0.3 to 0.6 GeV/$c$. This clear
separation has been used to calibrate the TPC $dE/dx$ without other
means of identification. It provides the fixed points for
extrapolating the $dE/dx$ function to higher momentum. In the thin
material (TPC gas) the Bichsel function has proved to be a very good
approximation for the $dE/dx$ curves and has been adopted by STAR as
a standard method of predicting the $dE/dx$ value for charged
hadrons in all momentum ranges~\cite{bichsel}.

The Landau function is an approximation which does not include
features related to atomic structure. With the Bichsel
function~\cite{bichsel}, a decrease of the relativistic rise of
$dE/dx$ with increasing segment length $x$ is seen and parameterized
empirically. This effect, gas multiplication gains and noise of TPC
electronics and pileup in high luminosity environment may make the
$dE/dx$ deviate from the Bichsel function. In the relativistic rise
region, the $dE/dx$ separations among $\pi^{\pm},K^{\pm}$ and
$p(\bar{p})$ are about 1-3$\sigma$ with the $dE/dx$ amplitude of
pions the highest and that of protons the lowest. Pions are the
dominant particle species for inclusive and jet hadrons, and they
shadow kaons and protons in the $dE/dx$ distribution. In a given
momentum slice, clear peak separations of these three hadrons are
not possible. This results in large systematic errors due to the
uncertainty of the $dE/dx$ positions. From the minimum-bias
triggered p+p collisions in the year 2003, the $p(\overline{p}$)
spectra were measured to \pt~$\lesssim$ 7 GeV/$c$ with significant
systematic errors due to the uncertainties in determining the mean
$dE/dx$ position for proton and kaon~\cite{starppPID}. Knowledge of
the precise $dE/dx$ positions for those hadrons is important to
understand the efficiencies of PID selection and to reduce the
systematic uncertainty in identified hadron yields. In order to
improve the particle identification at high \pt, we develop a method
to locate the $dE/dx$ positions for different hadrons with good
precision.

The p+p collision events with enhanced high-$p_T$ charged particles
used in this analysis were obtained from online jet trigger by the
BEMC with 0 $<$ $\eta$ $<$ 1 in the year 2005, and full azimuthal
coverage. $\Lambda\rightarrow p+\pi^{-}$
($\overline{\Lambda}\rightarrow\overline{p}+\pi^{+}$) and
$K_S^{0}\rightarrow\pi^{+}\pi^{-}$ are reconstructed by their decay
topology to identify their decay daughters -- charged pions and
protons. The identified electrons, pions and protons provide the
necessary distinct $dE/dx$ positions and widths as function of
$\beta\gamma$. The deviation from the prediction of the Bichsel
function was used to correct for the $dE/dx$ fit when extracting
pion and proton yields from the charged hadrons in an inclusive
hadron distribution or in a jet. The same method can be applied to
p+p, d+Au and A+A collisions in STAR.

Physics goals of the STAR Experiment at RHIC in recent (and future)
years~\cite{starwhitepaper} drive the need to operate the STAR TPC
at ever higher luminosity, and increase the ionization density in
the TPC gas~\cite{tpcDistortion}. The resulting ionic space charge
and grid leakage introduce field distortions in the detector, which
systematically shift the reconstructed momentum of positive and
negative particles in oppositive directions. The effect is expected
to grow as a function of $p_T$. STAR has developed a method for
correcting the track reconstruction due to space charge distortion.
Performance of the corrections can be assessed by examining the
distribution of signed DCA (Distance of Closest Approach of a
primary track to the collision vertex) as a function of luminosity.
In this paper, we use electrons and positrons dominantly from gamma
conversion to check and correct for potential residuals from the
distortion. We expect $e^{-}/e^{+}$ to be unity independent of $p_T$
since $\gamma\rightarrow e^{-}+e^{+}$ and a significant fraction of
leptons from heavy-flavor decays are also expected to be close to
unity~\cite{starcharm}. The high-statistics data set from BEMC
trigger is ideal for such a study.

\section{Experimental Setup and Data Analysis}
\subsection{Detectors and Datasets}
The Solenoidal Tracker at RHIC (STAR)~\cite{STAR} is a powerful
detector with a large and uniform acceptance capable of tracking
charged particle and providing particle identification (PID). Its
main tracking detector -- TPC~\cite{TPC} covers full azimuthal angle
( $\Delta\phi$ = 2$\pi$ ) and -1.3 $<\eta<$ 1.3 in pseudo-rapidity,
and provides identification of the charged hadrons by measuring
momentum and $dE/dx$ separation of the charged particles.  The STAR
TPC has 45 rows of pads with the first pad row at a radius of 60 cm
and last at 180 cm, providing a maximum of 45 sampled $dE/dx$ points
with 78 cm track length in a P10 gas (90\% Argon and 10\% $CH_4$)
while Track lengths of up to 110 cm are possible. In order to obtain
the $dE/dx$ position of protons at the relativistic rise range,
protons (anti-protons) are selected from $\Lambda$ ($\bar{\Lambda}$)
decays. Similarly, decay pions from $K^{0}_{S}$ can be used to get
$dE/dx$ and $dE/dx$ positions of pions for $0.2<p_T<3$ GeV/$c$. The
precise track geometry allows the particle trajectories to be
extrapolated to form a $V0$ away from the collision
vertex~\cite{Topology}. This provides $dE/dx$ information from
samples of pure protons and charged pions.

In addition, the sub-detector BEMC~\cite{BEMC} is also used to enhance
electron and positron yields and to confirm their identification. This
study used about 16.7 million p+p collision events recorded through
online High Tower (HT) and Jet Patch (JP) triggers in the BEMC. The HT
trigger condition required the energy of a single calorimeter tower
($\Delta\eta\times\Delta\phi$ = 0.05 $\times$ 0.05) to be at least 2.6
(HT1) or 3.6 (HT2) GeV, and JP trigger required the total energy of
one patch of towers ($\Delta\eta\times\Delta\phi$ = 1.0 $\times$ 1.0)
to exceed 4.5 (JP1) or 6.5 (JP2) GeV~\cite{Trigger}.

\subsection{$dE/dx$ distribution in the TPC at high $p_{T}$}
\begin{figure}
\centering
\includegraphics[width=0.45\textwidth]{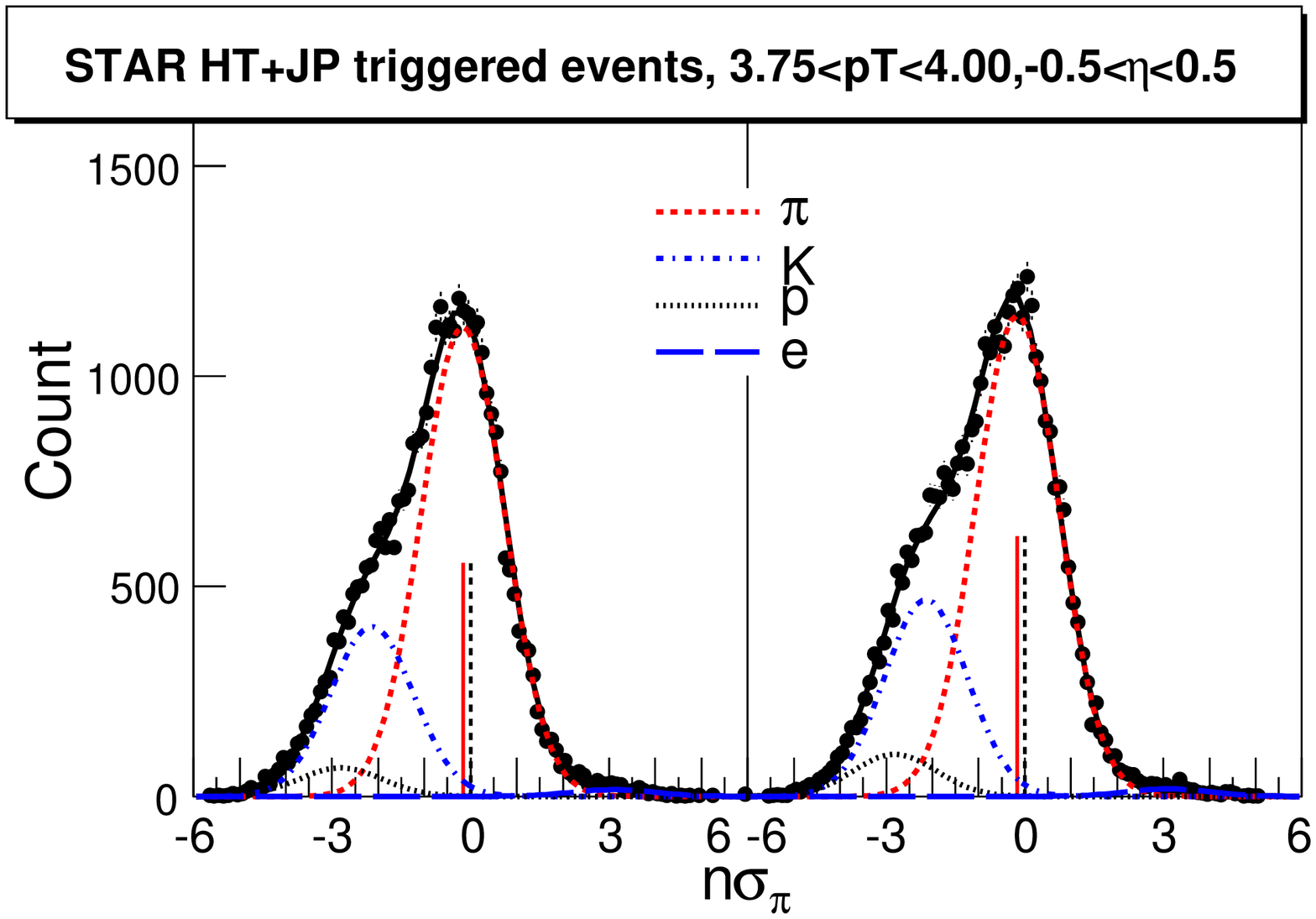}
\includegraphics[width=0.45\textwidth]{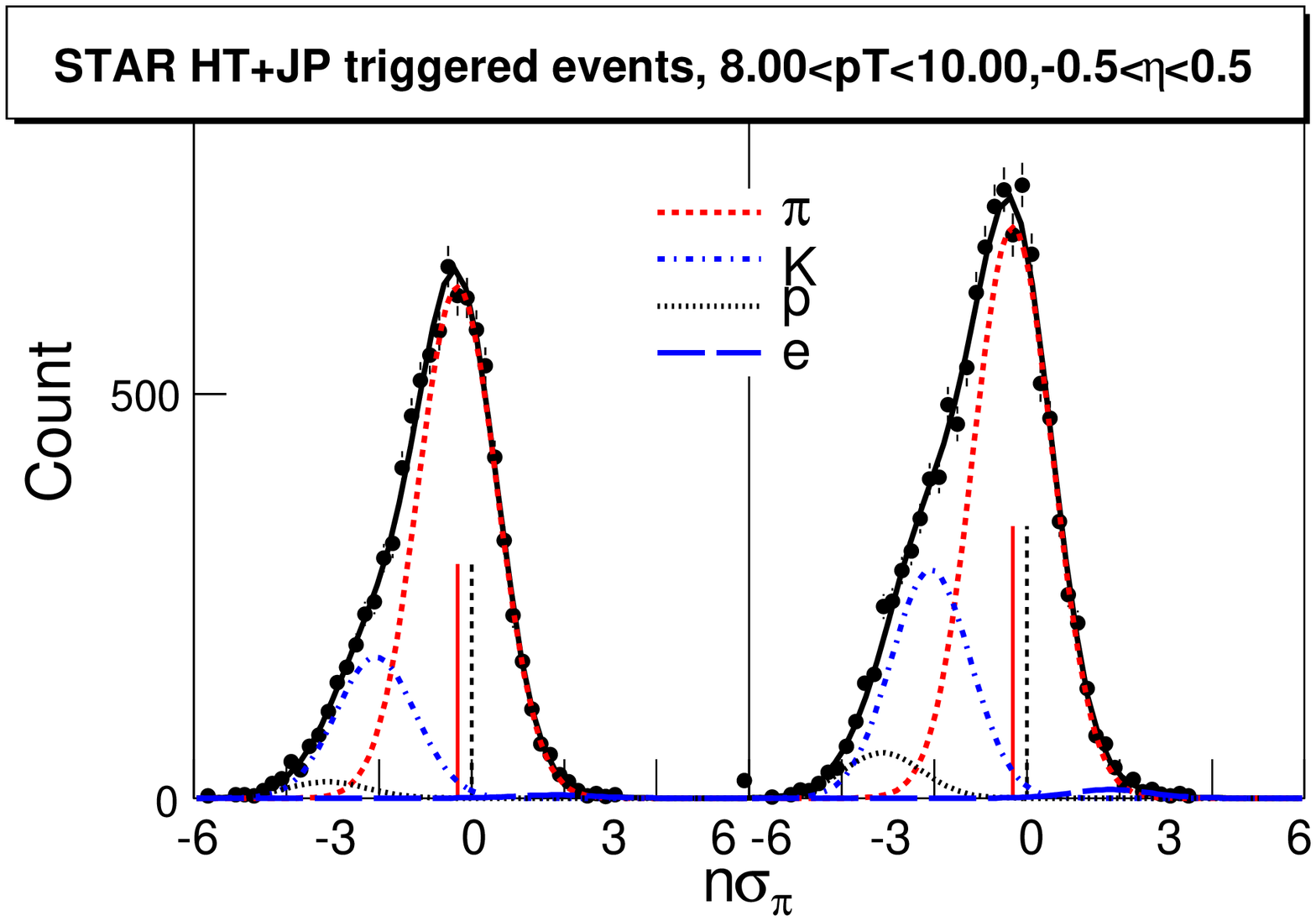}
\caption{$n\sigma_{\pi}^{h}$ distributions at 3.75$<$\pt$<$4.0 GeV/$c$
and 8.0$<$\pt$<$10.0 GeV/$c$ for the positive (left panel) and
negative (right panel) particles. The solid lines are the 8-Gaussian
function, which is a sum of the invidivual Gaussian functions from
pion (dashed line), kaon (dot-dashed line), proton (dotted line), and
electron (long-dash line). The solid vertical lines are the extracted
pion $dE/dx$ positions, while the dashed vertical lines are the
previously expected positions.}\label{nSigma}
\end{figure}

\begin{figure}
\centering
\includegraphics[width=0.6\textwidth]{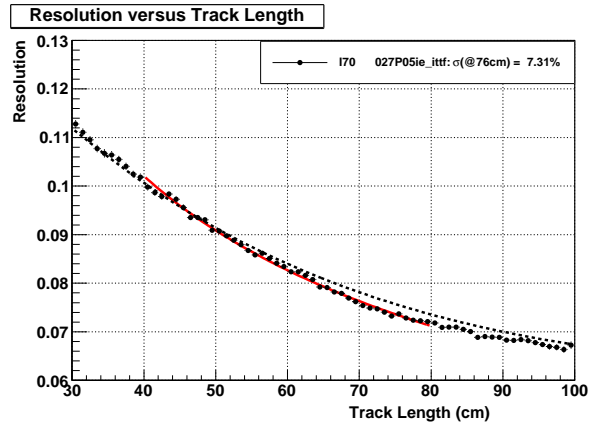}
\caption{dE/dx resolution ($\sigma_{\pi}$) as a function of track
length.  The red-solid line is a $\sigma_{0}/t^{0.52}$ power-law fit,
and the dashed line is a polynomail function upto 4th power in
$\ln{(t)}$.}\label{sigma2t}
\end{figure}

To formulate the $dE/dx$\footnote{$dE/dx$ is used to represent the
``track descriptor" C defined on $p.$170 of \cite{bichsel}
}distribution and its associated Bichsel function for PID, we need
to define a few terms. The normalized $dE/dx$ is defined as
$n\sigma_{X}^{Y} = \log[(dE/dx)_{Y}/B_{X}]/\sigma_{X}$, where X,Y
can represent $\pi^{\pm}$, $K^{\pm}$, $p(\overline{p}$) or
$e^{\pm}$; $B_{X}$ is the expected mean $dE/dx$ of particle X; and
$\sigma_{X}$ is the $\ln(dE/dx/B_X)$ resolution of the
TPC~\cite{Ming,lowptPID,lndEdx}. All the quantities are calculated
in a track-by-track basis. Fig.~\ref{nSigma} shows $n\sigma_{\pi}$
distribution of all the charged particles for 3.75$< $\pt$ < $4.0
GeV/$c$ and 8.0$< $\pt$ < $10.0 GeV/$c$ at $\mid\eta\mid<0.5$. A sum
of eight Gaussian functions is used to fit this distribution with
thirteen parameters in order to obtain the identified hadron yields.
Each Gaussian describes one $dE/dx$ distribution for a charged
particle species. The thirteen parameters are peak positions
relative to pion peak ( $n\sigma_{\pi}^{\pi}$,
$n\sigma_{\pi}^{K}$-$n\sigma_{\pi}^{\pi}$,
$n\sigma_{\pi}^{p}$-$n\sigma_{\pi}^{\pi}$,
$n\sigma_{\pi}^{e}$-$n\sigma_{\pi}^{\pi}$ ), eight yields for the
charged particles and one common Gaussian width.  With ideal
calibration, $n\sigma_{\pi}^{\pi}$ should be a normal Gaussian
distribution centered at zero and with a width of unity. The
normality of Gaussians depends on the precision of $\sigma_{X}$ in a
track-by-track basis. As presented in Ref.~\cite{bichsel} (section
6.2 and Fig. 21, 22, and 23), the $\sigma_{X}$ depends on the track
length ($t$) in the $dE/dx$ sample. An empirical parameterization
yields $\sigma_{\pi}=\sigma_{0}/t^{s}$ where
$0.45<s<0.55$~\cite{bichsel}. Fig.~\ref{sigma2t} shows the $dE/dx$
resolution as a function of track length for this specific run in
p+p collisions at center of mass energy of 200 GeV in year 2005. The
red solid curve is a power-law fit with $s=0.52$ in the range of
$40<t<80$ cm, where the majority of the tracks come from. A
polynomial function was also performed at larger range and was shown
as dashed line.

Fig.~\ref{nSigma} shows that the pion $dE/dx$ position is different
from the Bichsel function. This means that the $dE/dx$ calibration
is not perfect and it also implies that $dE/dx$ position of other
particles relative to that of the pions may be off from their
expected values. In order to improve the particle identification and
reduce the systematic uncertainty in identified particle yields
obtained from $dE/dx$~\cite{starPIDpapers,starppPID}, we study in
details the precise $dE/dx$ positions of all charged particles using
the enhanced electron by the BEMC, pure proton decayed from from
$\Lambda$ and pion decayed from \Ks~in the TPC. Once all the $dE/dx$
positions and widths for all the charged hadrons are obtained by
other means, we will be able to constrain better the Gaussian fits,
and understand the efficiency and contamination better in the case
of PID selections for other physics analysis.

\subsection{Electron identification by the BEMC}
\begin{figure}
\includegraphics[width=0.45\textwidth]{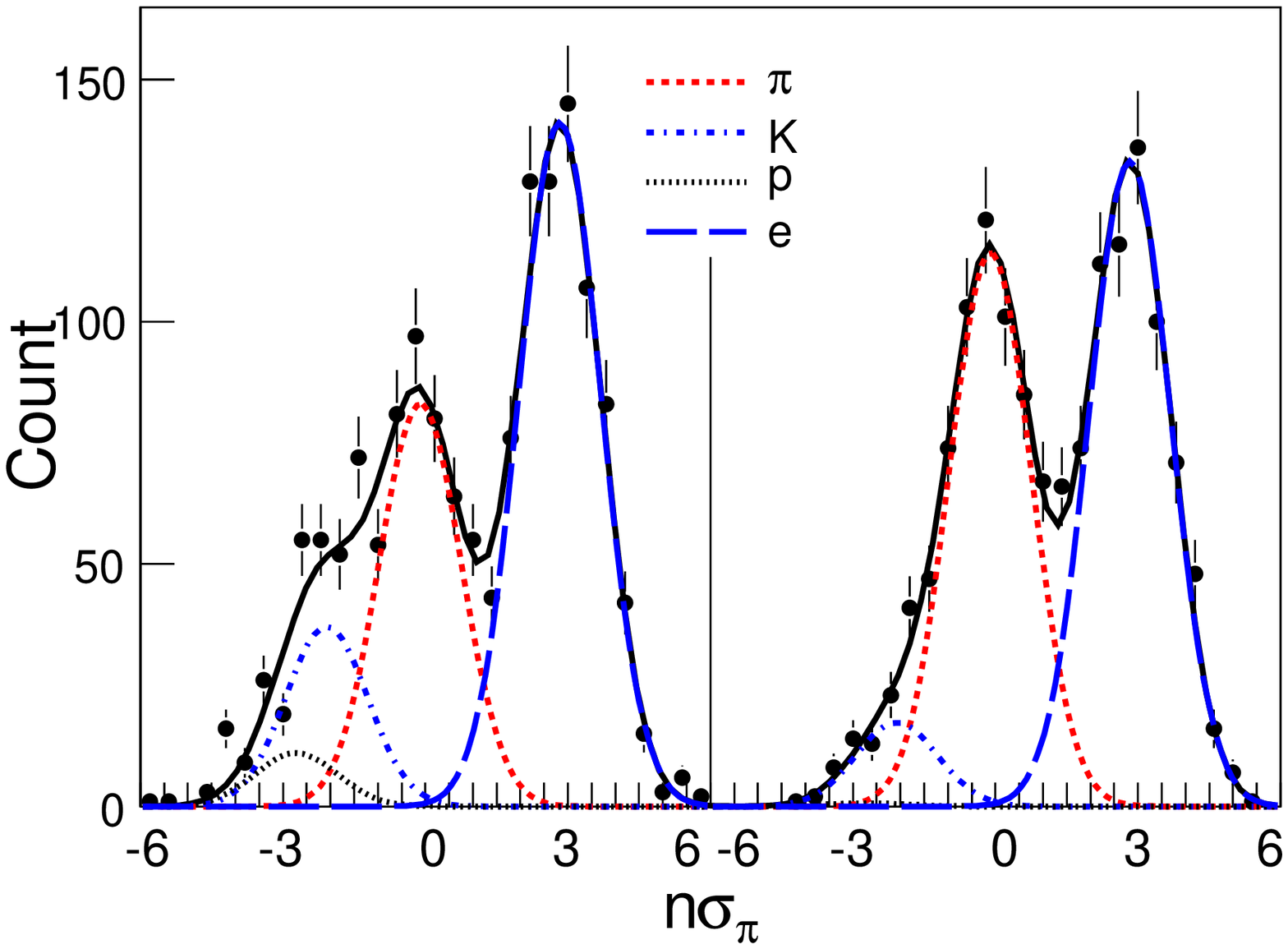}
\includegraphics[width=0.45\textwidth]{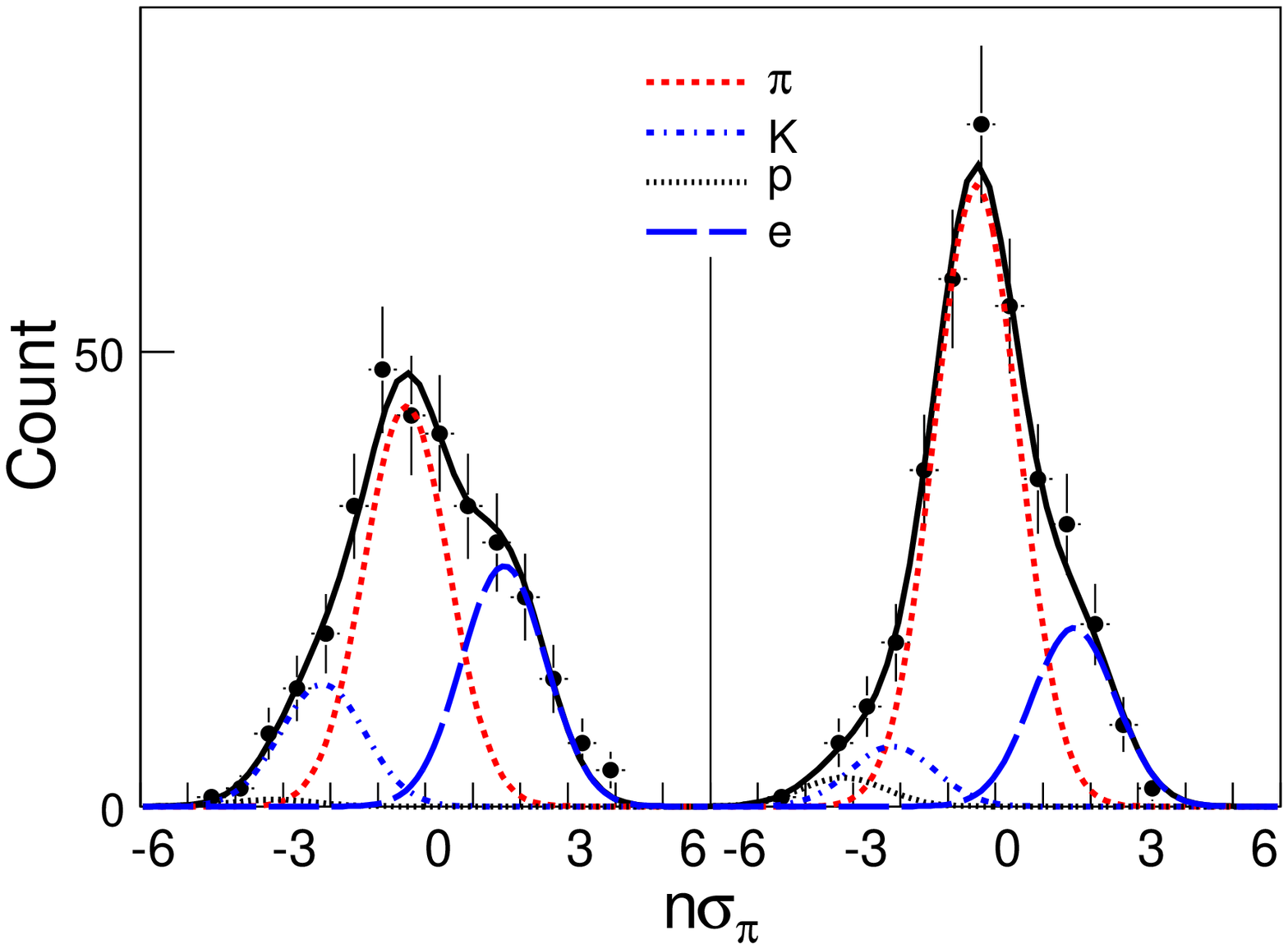}
\caption{$n\sigma_{\pi}^{h}$ distribution with high-tower trigger at
3.75$< $\pt$ < $4.0 GeV/$c$ (left panel) and 8.0$<$\pt$<$10.0
GeV/$c$ (right panel). Also shown is the result from an 8-Gaussian
fit. Both electron and positron yields are enhanced significantly by
the trigger.}\label{electron}
%\end{figure}
%\line(0,2){120}
%\begin{figure}
\end{figure}

Although electron $dE/dx$ is relatively far away from those of the
other charged particles, the electron yields are orders of
magnitudes lower than the yields for pions. In order to identify the
electron and obtain its $dE/dx$ position, a dataset with a special
trigger based on the energy deposited in the BEMC tower is used to
enhance the yield of electrons relative to other particles.
Additional hadron rejection is achieved from the shower shape and
position from a Shower Maximum Detector (SMD)~\cite{BEMC}, which is
located at $\sim$5.6 radiation lengths depth in the BEMC. A unique
feature of the SMD is its double layer design which makes it
possible to reconstruct the shower as two-dimensional image, so that
it can provide fine spatial resolution in $\phi$ and $\eta$
directions and reject hadrons according to the different shower
shape between hadrons and electrons. We require $p/E$ to be 0.3$<$
$p/E$ $<$1.5 where $p$ is track momentum in the TPC and $E$ is the
energy of the BEMC tower, and the shower hits of the SMD in $\eta$
and $\phi$ direction to be $n_{\eta}\geq$ 2 and $\phi$ direction
$n_{\phi}\geq$ 2, respectively. The different positions between hit
and track projection in $\phi$ and $z$ directions are restricted to
be $\mid\phi_{dist}\mid\leq$ 0.01 rad and $\mid z_{dist}\mid\leq2$
cm~\cite{BEMC}. The $n\sigma_{\pi}^{h}$ distribution for track after
these cuts are shown on Fig.~\ref{electron} for 3.75 $<$ \pt $<$ 4.0
GeV/$c$ and at much higher \pt~range. With about 1.5-3$\sigma$
separation between electron (positron) and other hadrons, electron
position and yields could be obtained from the eight-Gaussian
function as above. From the same fit, we can also obtain the $dE/dx$
positions for pions, which are labeled as ``pion with EMC'' for
later use.

\begin{figure}
\center{\includegraphics[width=0.6\textwidth]{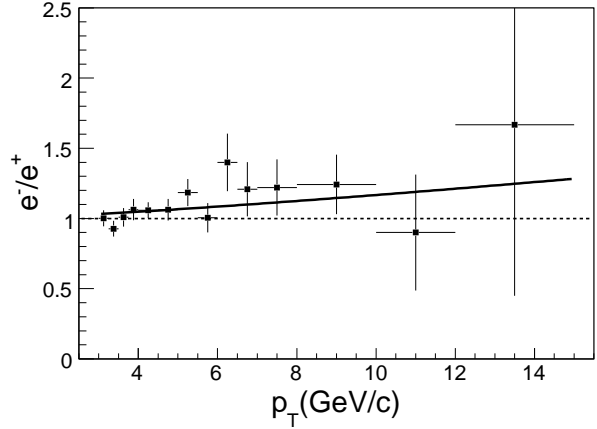}}
\caption{ The ratio of $e^{-}/e^{+}$ as a function of \pt. The curve
is a power-law fit described in the text.}\label{empRatio}
\end{figure}

To correct for and to assess the systematic errors due to the
residual momentum distortion, we use $e^-/e^+$ ratio as a function
of $p_T$. Fig.~\ref{empRatio} shows the ratio as a function of $p_T$
obtained from the BEMC triggered data described as above. If the
electron and positron yields are a modified power-law function
($f(p_T)\propto(p_0+p_T)^{-n}$) without any distortion, the
distortion due to space charge in the TPC shifts all negatively
charged tracks from $p_T$ value to a higher $p_T+A\times p_T^{2}$
while it shifts all positively charged tracks from $p_T$ value to a
lower $p_T-A\times p_T^{2}$~\cite{tpcDistortion}. The correction
form is according to the effect of sagitta displacement due to space
charge on momentum distortion~\cite{PDGsagitta}. Data points are
fitted by the following function $f(p_{T})
=(\frac{2.67+p_{T}+A*p_{T}^{2}}{2.67+p_{T}-A*p_{T}^{2}})^{11.4}$,
where $A*p_{T}^{2}$ means $\Delta p_{T}$ affected by charge
distortion, and $p_0=2.67$ and $n$ = 11.4 are parameters obtained
from the inclusive electron spectra~\cite{starcharm}. The $p_{T}$
dependence of ratios indicates that the momenta of the charged
particles obtained from the TPC tracking still systematically
shifted away from their true value due to the space charge
distortion. We note that the obtained distortion characterized by
parameter {\bf $A$} is only about $2\sigma$ from zero and this
results in about $1.3\%$ momentum shift for a single particle track
at $p_T$ = 15 GeV/$c$.

\subsection{Proton and pion from $V0$ reconstructed in the TPC  }

\begin{figure}
\begin{minipage}[t]{6.8cm}
\center{\includegraphics[width=0.9\textwidth]{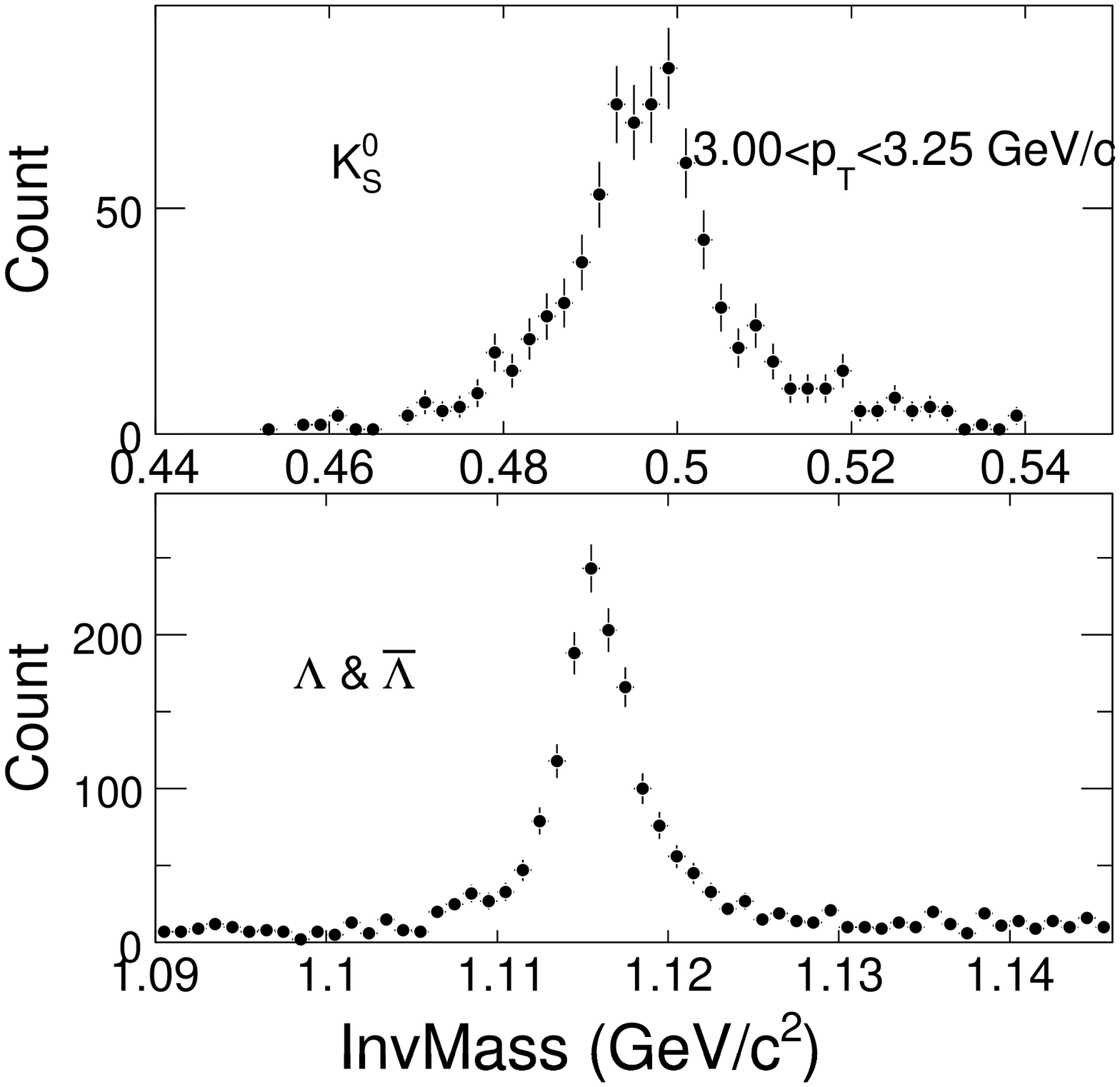}}
\caption{Invariant mass distributions of \Ks (upper panel) and
$\Lambda$ (lower panel) with at least one daughter particle ($\pi$
or proton) at 3.0$<$\pt$<$3.25 GeV/$c$.}\label{LambdaKshort}
\end{minipage}
\hspace{0.05in}
\begin{minipage}[t]{6.8cm}
\center{\includegraphics[width=0.9\textwidth]{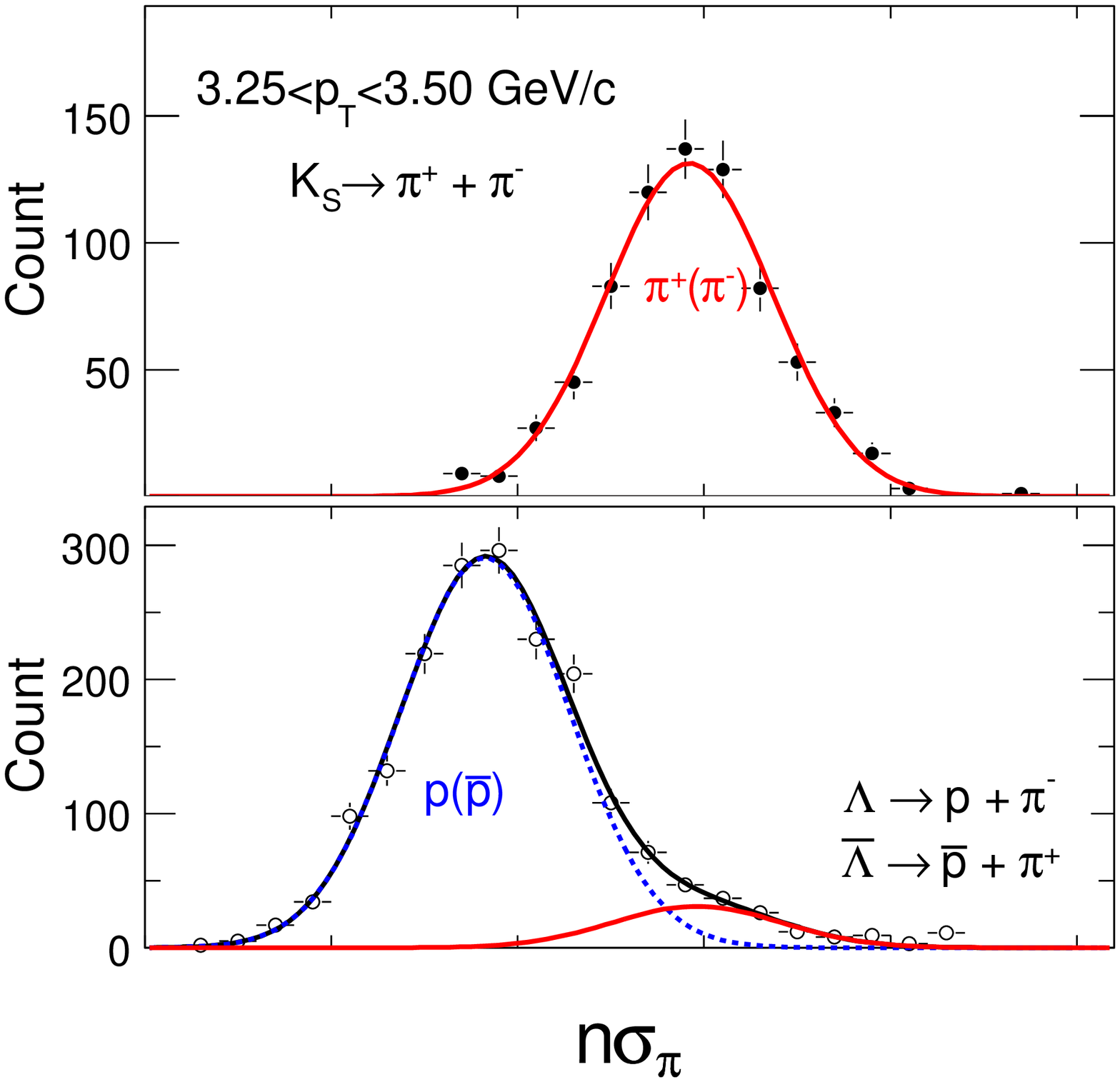}}
\caption{$n\sigma_{\pi}^{h}$ distributions of pions from \Ks (upper)
and protons (lower) from $\Lambda$. The solid line is for pions, and
the dashed one for protons.}\label{protonpion}
\end{minipage}
\end{figure}

\begin{figure}
\begin{minipage}[t]{6.8cm}
\center{\includegraphics[width=0.9\textwidth]{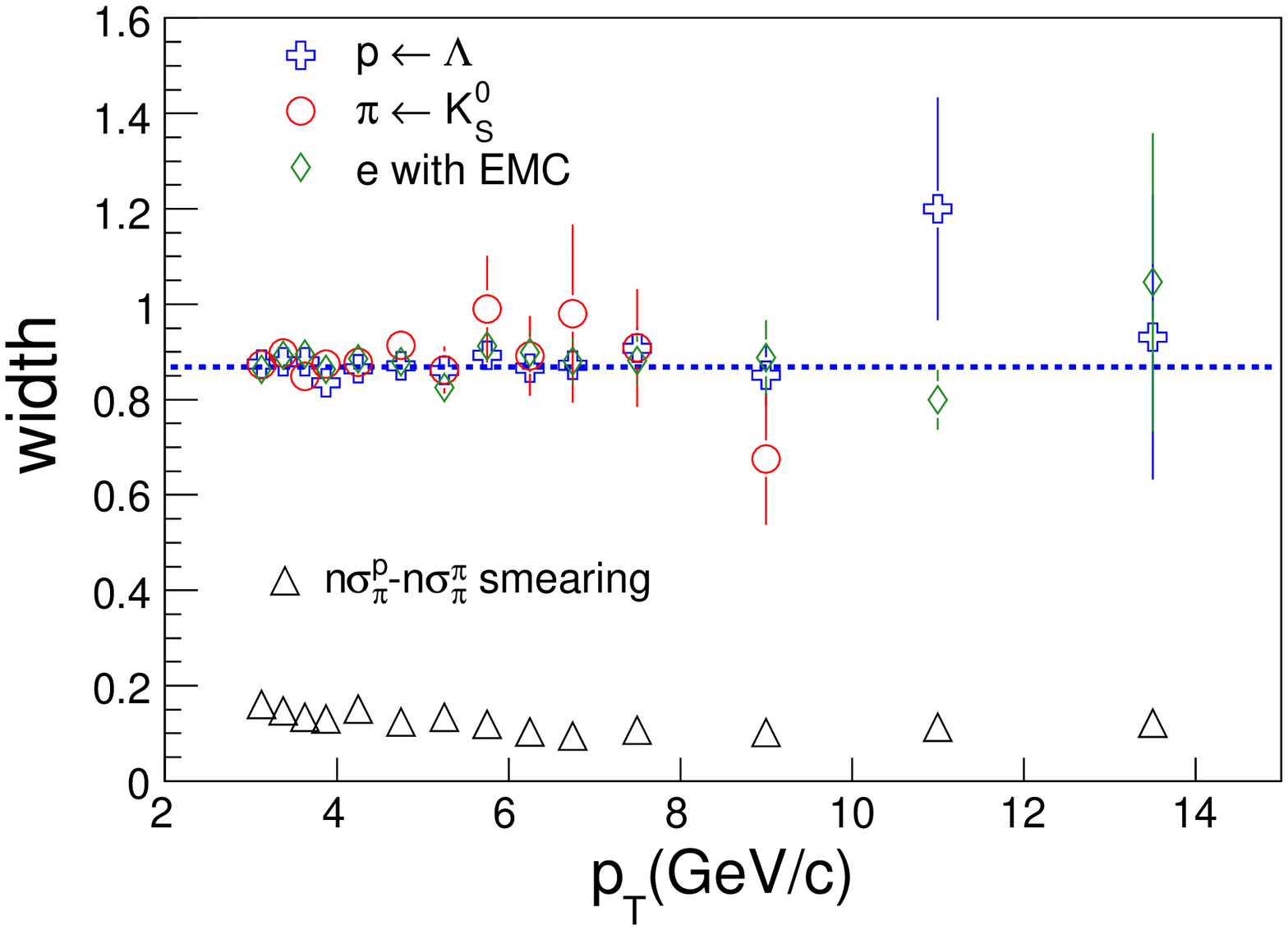}}%widthPlot4.eps}}
\caption{\pt~dependence of $dE/dx$ $n\sigma_{\pi}^{h}$ width for
proton from \La decay (crosses), for pion from \Ks decay (circles),
and for electron enhanced by the EMC (diamond). The open triangle is
the smearing between proton and pion dE/dx peak positions due to the
variation of the track quality ($\sigma^{p}_{\pi}$). }\label{width}
\end{minipage}
\hspace{0.05in}
\begin{minipage}[t]{6.8cm}
\center{\includegraphics[width=0.9\textwidth]{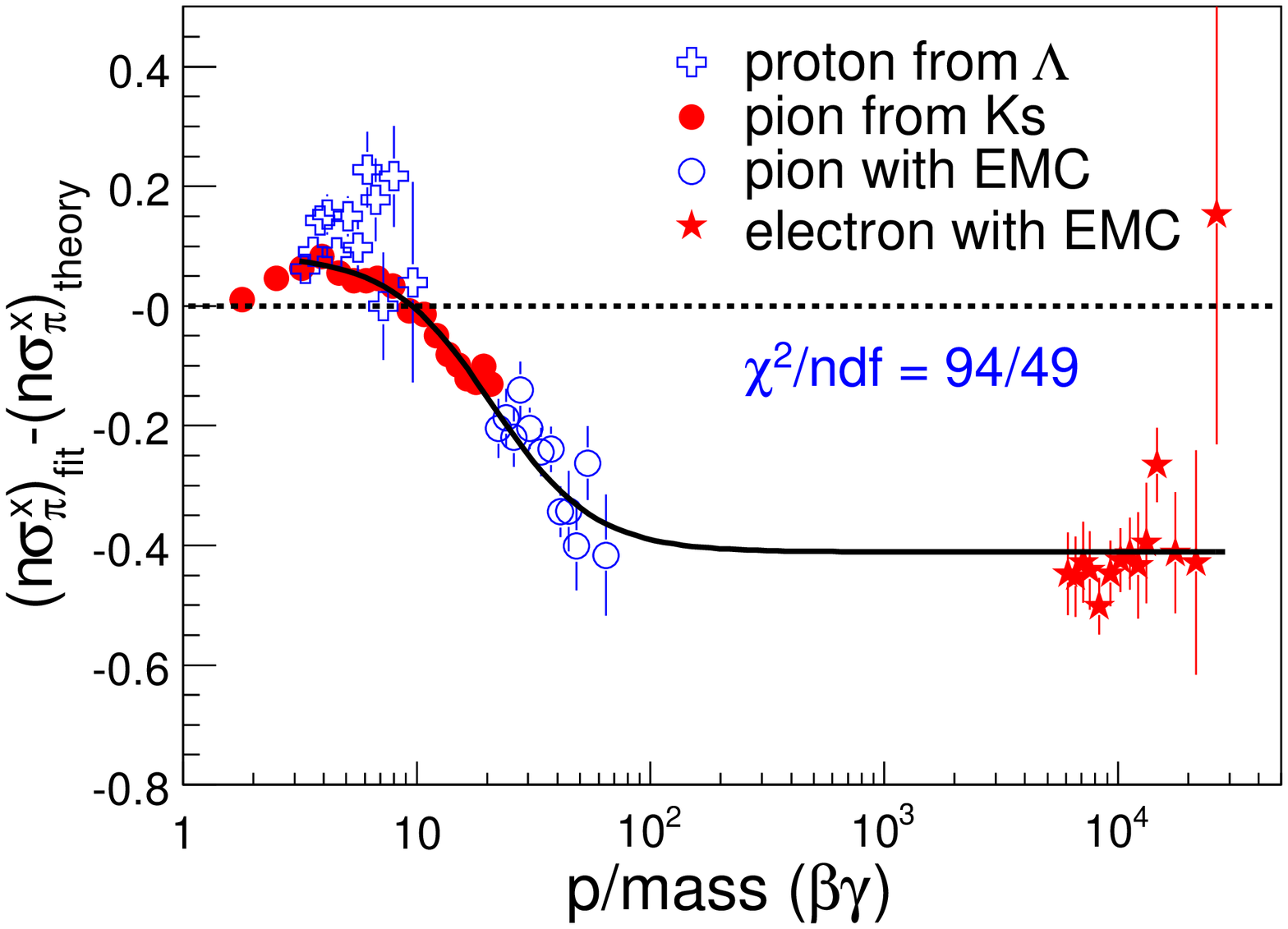}}
\caption{The $dE/dx$ deviation as a function of $\beta\gamma$. The
crosses are for protons from $\Lambda$, the filled dots are for
pions from \Ks, and open circles and stars are for pions and
electrons from HT trigger, respectively.}\label{positionfitp}
\end{minipage}
\end{figure}

In order to obtain the $dE/dx$ position of proton
($n\sigma_{\pi}^{p}$) for $\beta\gamma>4$, protons (anti-protons)
are selected from $\Lambda$ ($\overline{\Lambda}$) through
$\Lambda\rightarrow p+\pi^{-}$ ($\overline{\Lambda}\rightarrow
\overline{p}+\pi^{+}$) decays, because it is difficult to get
$dE/dx$ position of proton by $h^{+}-h^{-}$ \cite{Ming} with this
data sample due to low statistics and small difference of yields of
proton and anti-proton. At the same time, pions decayed from \Ks
through $K^{0}_{S}\rightarrow \pi^{+}+\pi^{-}$ decay can be used to
get $dE/dx$ and $dE/dx$ positions of pion ($n\sigma_{\pi}^{\pi}$) at
0.2$<$\pt$<$3 GeV/$c$. First, \Ks and $\Lambda$ are selected by
topological cuts on a secondary vertex~\cite{Topology} according to
long decay length of \Ks ($c\tau$=2.6 cm) and $\Lambda$
($c\tau$=7.89 cm).  Fig.~\ref{LambdaKshort} shows the invariant mass
distribution for \Ks (top panel) and $\Lambda$ (bottom panel). Pure
\Ks and $\Lambda$ ($\overline{\Lambda}$) are selected via their
invariant mass cuts, 0.485$<$M(\Ks)$<$0.505 GeV/$c^{2}$ and
1.112$<M(\Lambda)<1.12$ GeV/$c^{2}$, and their daughter particles
($\pi$,$p(\overline{p}$)) with high purity are obtained to derive
$n\sigma_{\pi}^{p}$ and $n\sigma_{\pi}^{\pi}$. Fig.~\ref{protonpion}
shows $n\sigma_{\pi}^{h}$ distribution of pions decayed from \Ks
(upper panel) fitted by a Gaussian function and protons decayed from
$\Lambda$ (lower panel) fitted by a 2-Gaussian function. The protons
from $\Lambda$ decay have higher background (signal-to-background
ratio = 9:1) and a second Gaussian representing the pion
contamination is necessary. Meanwhile, the $p_{T}$ dependence of the
$dE/dx$ width for protons and pions from the fits are shown in
Fig.~\ref{width}, which also shows $dE/dx$ width for electron. The
width is consistently smaller than unity ($0.868\pm0.004$). This
means that the $dE/dx$ resolution is about 13\% better than the
prediction and the separations among particles are better than what
we expected. This may be due to the run to run variation
(luminosity, beam background, etc.), and our calibration only
sampled a small fraction of data($\sim5\%$). The open triangles in
Fig.~\ref{width} shows the sole effect of the smearing of the
$dE/dx$ peak positions between proton and pion due to the variation
of the track quality ($\sigma^{p}_{\pi}$). This effect
($<0.2\sigma$), compared quadratically to the dE/dx width,
contributes $<4\%$ to the final $dE/dx$ width.

\subsection{$dE/dx$ deviation vs $\beta\gamma$}

With identified pion, proton and electron mentioned above, the
experimental results on the deviation of the normalized $dE/dx$
($n\sigma_{\pi}^{h}$) relative to the Bichsel theoretical values, as
a function of $\beta\gamma$ are shown on Fig.~\ref{positionfitp}.
Since there is almost no particle species dependence of $dE/dx$, we
can describe it with a function of $f(x) = A + \frac{B}{C+x^{2}}$.
The fit parameters are listed in Tab.~\ref{enhancefactor}. With
this, we can determine the $dE/dx$ positions and widths of any given
charged particles to better than $<0.1\sigma$ or $<1\%$.

\begin{figure}
\begin{minipage}[t]{6.8cm}
\center{\includegraphics[width=0.9\textwidth]{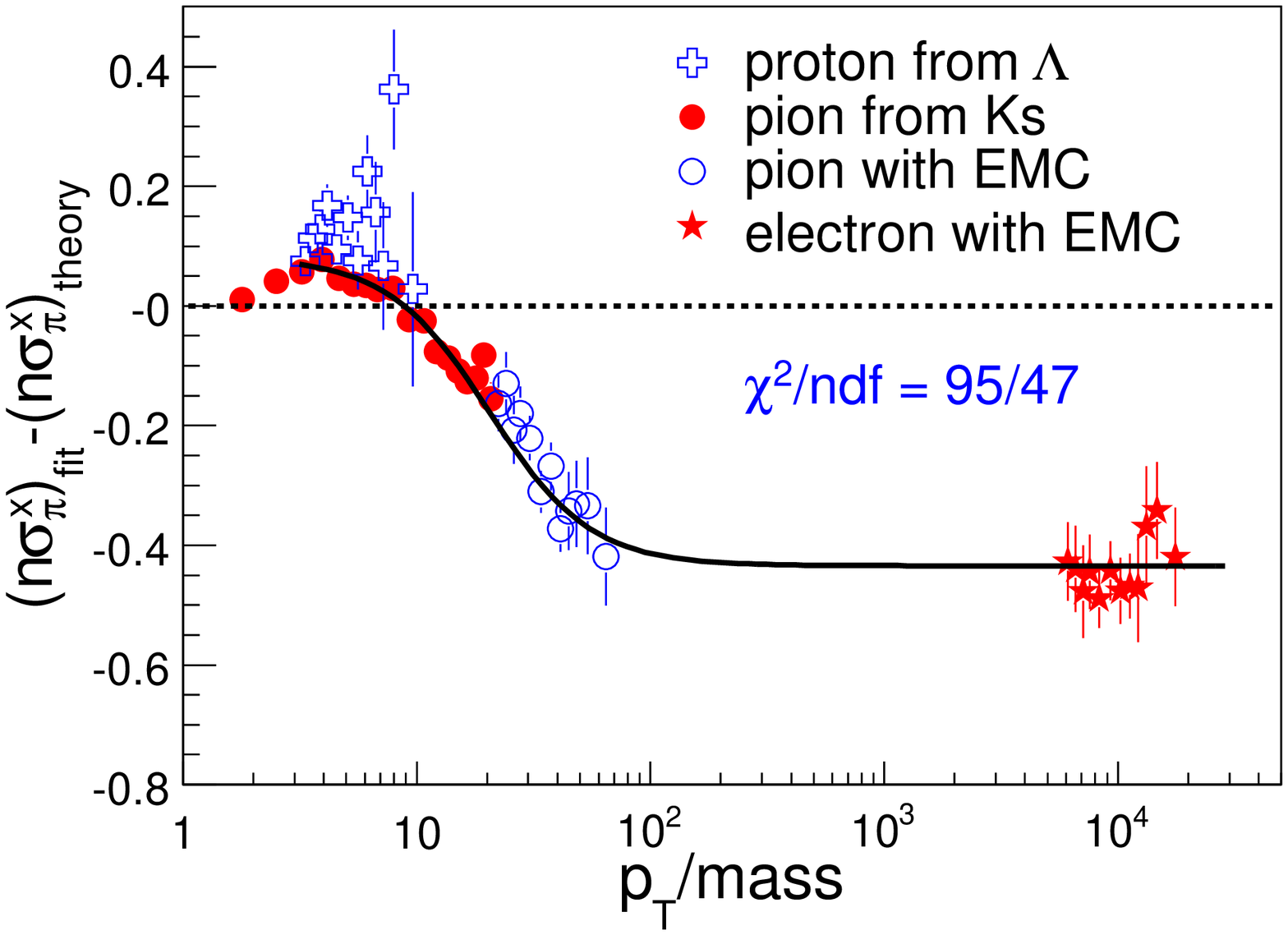}}
\caption{The $dE/dx$ deviation as function of $p_T$/mass (similar to
Fig.~\ref{positionfitp}).}\label{positionfitpt}
\end{minipage}
\hspace{0.05in}
\begin{minipage}[t]{6.8cm}
\center{\includegraphics[width=0.9\textwidth]{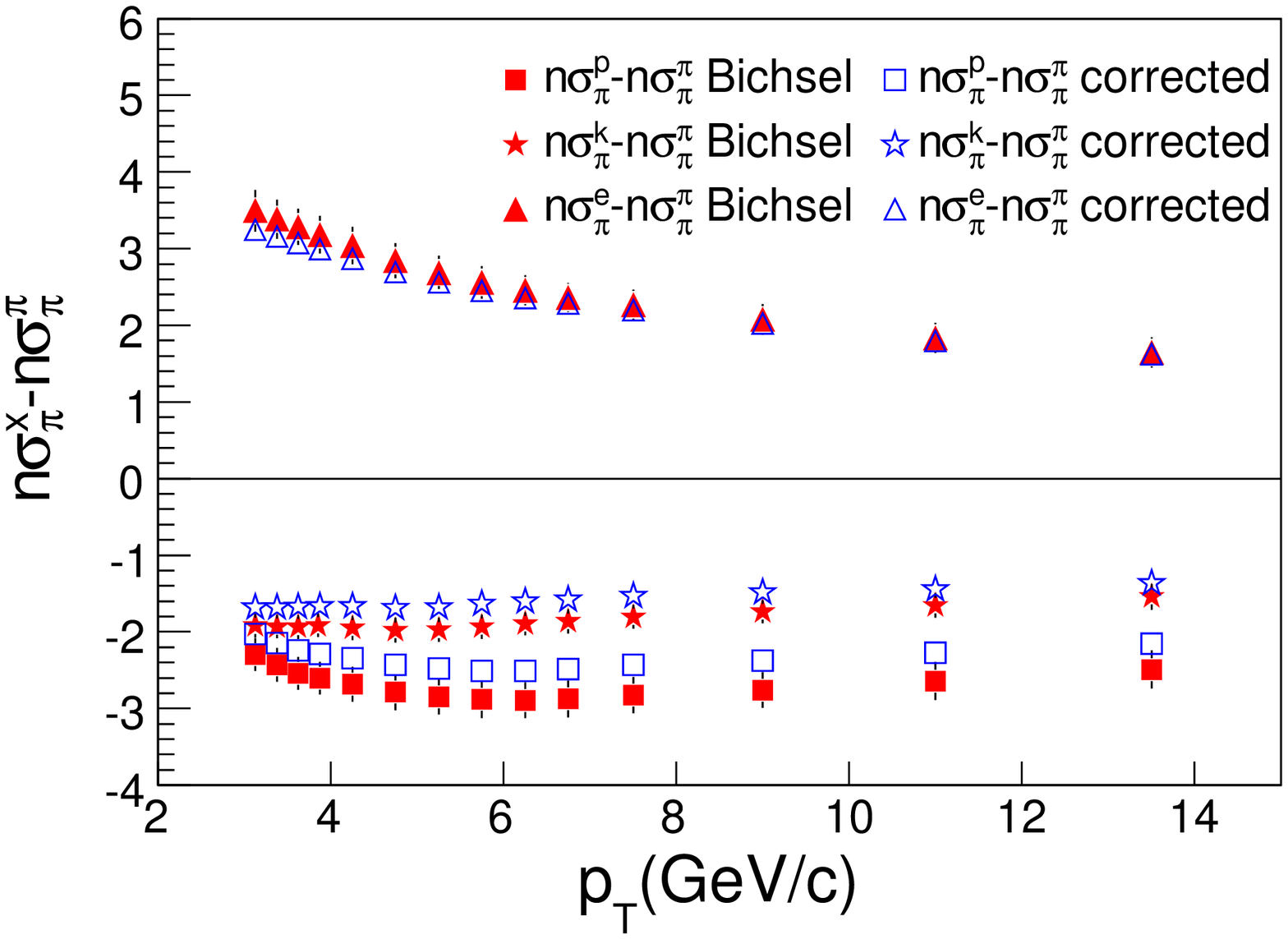}}
\caption{The relative $dE/dx$ peak positions of $n\sigma_{\pi}^{K}$,
$n\sigma_{\pi}^{p}$, $n\sigma_{\pi}^{e}$ as a function of $p_T$. The
solid dots depict theoretical values, and open ones are re-calibrated
results.}\label{nSigmaDifference}
\end{minipage}
\end{figure}

There are two ways to correct for this effect in the data. One can
attempt to understand the origin of this deviation and correct for
the effect at the hit level (the amplitudes of the ionization signal
in each pad and row). This requires re-processing of the hits and
reconstructing tracks from scratch. In order to take advantage of
the existing compressed dataset with tracking information only, we
apply the corrections to the $dE/dx$ Gaussian function for each
particle species without modifying the $dE/dx$ itself. This
empirical afterburner is applied in each $p_T$ and rapidity bin to
directly extract particle yields required by the physics analyses.
Fig.~\ref{positionfitpt} shows \pt/mass dependence of the normalized
$dE/dx$ deviation, which is fitted by the same function as for the
case of $p$/mass. The parameters from the function are shown in
Tab.~\ref{enhancefactor}. With the corrected deviation, differences
of $dE/dx$ between pion and other charged particles
($n\sigma_{\pi}^{K}-n\sigma_{\pi}^{\pi}$,
$n\sigma_{\pi}^{p}-n\sigma_{\pi}^{\pi}$ and
$n\sigma_{\pi}^{e}-n\sigma_{\pi}^{\pi}$) are calculated and compared
with theoretical values and shown in Fig.~\ref{nSigmaDifference}.
Clear offsets are depicted in Fig.~\ref{nSigmaDifference}.

\section{Summary}
The $dE/dx$ positions and widths of charged particles have been
precisely determined using BEMC triggered events with enhanced
electron content and pure samples of pions and protons from K0s and
Lambda decays respectively. Their deviations relative to theoretical
values versus $\beta\gamma$ and $p_T$/mass are described by the
function $f(x)=A+\frac{B}{C+x^{2}}$ very well. With this method,
$dE/dx$ positions of charged particles are re-calibrated to be
better than $0.1\sigma$. The particle identification of charged
hadrons is thus improved, and the uncertainty is reduced
significantly. These are important steps toward fulfilling the
physics goals of the STAR experiment at RHIC in the future.

\section*{Acknowledgments}
We thank the STAR Collaboration, the RHIC Operations Group and RCF
at BNL, and the NERSC Center at LBNL for their support. This work
was supported in part by the Offices of NP and HEP within the U.S.
DOE Office of Science; Authors Yichun Xu and Zebo Tang are
supported in part by NSFC 10475071, National Natural Science
Foundation of China  under Grant No. 10610286 (10610285) and
Knowledge Innovation Project of Chinese Academy of Sciences under
Grant No. KJCX2-YW-A14. One of us (Lijuan Ruan) would like to
thank the Battelle Memorial Institute and Stony Brook University
for support in the form of the Gertrude and Maurice Goldhaber
Distinguished Fellowship.

% \appendix

% \section{}
% \label{}

\newpage

\newpage

\newpage

\newpage

\newpage

\newpage

\newpage
\begin{table}
\begin{center}
\caption{Fit parameters from momentum and $p_{T}$ dependence of
$n\sigma_{\pi}^{x}$ with the function $f(x) =
A+\frac{B}{C+x^{2}}$.}\label{enhancefactor}
\begin{tabular}{|c|c|c|c|c|p{6.0cm}}
  \hline \hline
     parameters  & $\chi^{2}/ndf$   & A  & B & C \\
  \hline
    p dependence &95/49   & -0.423 $\pm$ 0.015  &235 $\pm$ 23 &464 $\pm$ 37\\
  \hline
    $p_{T}$ dependence  &94/48  & -0.443 $\pm$ 0.015 &234 $\pm$ 23 &444 $\pm$ 35 \\
  \hline \hline
\end{tabular}
\end{center}
\end{table}

\end{document}